# Nonequilibrium Kinetic Study of Sintering of Dispersed Nanoparticles

Vyacheslav Gorshkov,[a] Vasily Kuzmenko,[a] and Vladimir Privman[b,*]

[a] National Technical University of Ukraine — KPI, 37 Peremogy Avenue, Building 7, Kiev 03056, Ukraine

[b] Center for Advanced Materials Processing, Department of Physics, Clarkson University, Potsdam, NY 13699, USA

**Abstract**

Kinetic Monte Carlo approach is developed to study aspects of sintering of dispersed nanoparticles of bimodal size distributions. We explore mechanisms of neck development when sintering is initiated at elevated temperatures for nanosize crystalline surfaces of particles of different sizes. Specifically, the role of smaller particles fitting between larger particles, on the sintering of the latter is considered. Formation of stable necks bridging particles at the nanoscale was found to be governed by layering or clustering mechanisms at the facing surfaces, with clustering leading to a much faster formation of the bridging structure. Temperature, particle sizes and local arrangement, as well as other geometrical factors were found to have a profound effect on sintering mediated by a smaller particle placed in a void between larger particles.

___

[*] Corresponding author: e-mail privman@clarkson.edu; phone +1-315-268-3891

# 1. Introduction

Sintering is an important technological process which has attracted substantial efforts both in numerous experimental investigations, many of which very recent,[1-8] and in modeling by various techniques[9-21] ranging from continuum theories to numerical approaches. Various aspects of sintering have been studied, including the overall structure of the sintered materials and more local properties of neck formation and particle merging. Multiscale approaches to modeling[13,16,17] involve continuum/finite-element studies,[14,15] as well as more atomistic kinetic Monte Carlo[11,16,17] and molecular dynamics[18] methodology. Recently, there has been interesting new developments[4] utilizing the new capabilities of synthesizing nanoparticles of well-defined size, shape, and surface morphology. Incorporation of nanoparticles can improve the resulting connectivity, and thus, conductance,[4] in sintering of metal particles initially dispersed[2-6,8,22-26] in pastes and similar materials typically containing other components and additives. The particles can for example be mixtures of micron size polycrystalline colloids or smaller crystalline nanoparticles, of various predesigned sizes and shapes, from spherical to flake-like. Better connectivity of the metal in the resulting products (e.g., films), leading to improved conductance, can be obtained by the selection of parameters of the initial dispersion, for instance, adding smaller particles to securely fit in the voids between larger particles.[4] The smaller particles can then be expected to facilitate the formation of contacts (necks) between the larger particles.

Mechanical properties, density, and most recently conductance of materials prepared by utilizing bi-modal or more complicated particle distributions and mixing conditions before sintering, have been studied experimentally and theoretically for some time.[4,12,19,21,22,27,28] Effects of nano-size particle feature dimensions, such as sizes of their faces that are in near contact and form necks on sintering, as well as other "geometrical" considerations are, however, not obvious and require theoretical verification. Furthermore, it would be useful to have a numerical approach which is versatile enough to allow a certain degree of optimization, for example, devising a temperature-control protocol for best results. For sintering, temperature has to be elevated. However, criteria should be balanced such as that the maximum temperature be limited to avoid damage to other components of the sintered material, or the duration of the peak heating should be made shorter to lower costs. Other considerations can also be important, including for instance those related to the desired final structure of the sintered film and sometimes even for inhibiting[29] nanoparticle sintering when it interferes with other processes such as encapsulation.

Sintering results[21,23-33] from a competition of several kinetic processes: transport of matter, on-surface restructuring, and detachment/reattachment, which when combined, typically generate nonequilibrium dynamics. Here we explore and report the first results of possible applications of a recently developed kinetic Monte Carlo (MC) approach which incorporates such processes in a formulation suitable for describing the evolution of the surface and shape morphological features in sintering staring from preformed nanocrystals. We aim at considering the neck-formation scenarios in sintering starting with certain configurations of particles, focusing on the kinetics of the constituent atoms, ions or molecules, to be termed "atoms" for brevity. The model was recently developed[34] for growth of nanoparticles of well-defined shapes. It was also applied to formation of on-surface structures of interest for catalysis.[35,36]

Sintering has been studied and modeled for a long time. However, as emphasized in a survey of modeling of sintering processes (which also provides a comprehensive earlier literature citation list),[33] theoretical approaches have a limited predictive power at the quantitative level. Indeed, sintering is complicated and requires a multi-scale description[11,16,17] that none of the existing approaches can offer. Therefore, models of sintering[9-21,30-33,36-39] focus on particular features of the process and aim at a qualitative or at best semi-quantitative understanding of the process. Here we adopt the model which was devised[34] to apply in the regime in which nanoparticle shapes would emerge were they interacting with the supply of diffusing solute matter "atoms" in the surrounding medium. Well-defined particle shapes emerge at this stage of the kinetics for crystalline nanoparticles, after the formation of the initial core but before the onset of the large-scale growth that can destabilize the particle.

For sintering, the model is modified to consider the situation when more than one particle are closely packed and, rather than grow when matter is supplied during their synthesis, as studied earlier,[34] here they instead evolve kinetically by exchanging and competing for the diffusing "atoms" in the environment, including the transport of matter at their own surfaces.

Thus, the diffusing "atoms" in this case are not externally supplied to drive the growth, but can rather result from detachment off the initial particles. In the next, Theoretical section we survey some of the features of this approach as far as single (isolated) particle kinetics is concerned, as well as describe the model approach to study the situation when closely positioned particles affect each other's kinetics. Our Results and Discussion section reports studies of various geometries of sintering of closely positioned nanoparticles, including neck formation mechanisms at their initially nearly-touching faces. We explore the effect of sizes, temperature, and certain geometrical factors on sintering involving a smaller particle placed in a void between larger particles. In the Conclusion section we address the regime of applicability of the present numerical-simulation approach in the framework of studies of various stages and additional features of the sintering process, notably, volume shrinking of the emerging connected structure.

## 2. Theoretical Section

### 2.1. Outline of the Modeling Approach

Let us first outline, in this subsection, the approach used, based on earlier work on nanoparticle and nanostructure shape selection.[34-36] An important finding[34,40,41] has been that "persistency" can be a driving mechanism in the emergence and maintenance of well-defined shapes in nonequilibrium growth of nanoparticles and nanostructures. Here "nonequilibrium" is defined with respect to the overall rate of the restructuring/growth of the particle's surface, i.e., that the restructuring and matter transport processes are not fast enough to yield global thermal equilibration of the shape. Restructuring includes on-surface motion of atoms, as well as their detachment and attachment. The property termed imperfect-oriented attachment[40-45] has been identified as persistency in successive nanocrystal binding events leading to the formation of uniform short chains of aggregated nanoparticles. It has been established that persistency can also mediate growth of other shapes,[34-36,40,41,45] for a certain range of the particle and feature sizes. Nanosize particles and structures in fast-growth conditions are simply not sufficiently large (do not contain enough constituent atoms) to develop large internal defects and unstable surface features that result in the formation of whiskers and/or the "dendritic instabilities" of growing side branches, then branches-on-branches, etc. Such processes could distort a shape with approximately crystalline faces to cause it to evolve into a random/fractal or snowflake like morphology.[46,47]

The statements in the above paragraph apply to growth/surface kinetics of a single particle in an environment of abundant diffusing building-block atoms. Before considering more than one particle and their effect on each other, we outline the model[34] for several processes and their competition, which together result in the single-particle morphology and feature-shape selection. We assume diffusional transport and attachments of atoms to the evolving/growing particle surface. Atoms can also detach and reattach. They can move/hop/roll on the surface, according to thermal-like rules, but not fast enough to yield structure-wide thermalization on the time scales of the transport of diffusing matter to/from the surface. Diffusional transport without much restructuring processes would yield a fractal structure. Indeed, when the on-surface relaxation processes occur on time scales larger that than those of the formation of additional layers due to diffusional supply of matter, unstable, irregular-shape growth is expected. In the opposite regime, isolated particles would assume thermal-equilibrium Wulff shapes.[48-50]

We consider the practically important nonequilibrium regime when the two time scales are comparable. A uniformly proportioned (isomeric) shape selection for nanoparticles in this "noneqilibrium" regime has been explored.[34] The emergence of well-defined nanocrystal shapes for isolated particle growth has been reproduced by the kinetic MC approach,[34] which is also adopted here, for the standard crystalline symmetries, consistent with experimental nanoparticle shapes for metals. The model includes the kinetics of the atoms' hopping on the surface and detachment/reattachment, according to thermal-type, over free-energy barriers Boltzmann factor rules. The diffusion of atoms occurs in the three-dimensional space. However, atom attachment is only allowed "registered" with the underlying lattice of the initial structure(s). This prevents the growing structures from developing "macroscopic" (particle-wide/structure-wide) defects, which has been the property identified[34] as important for well-defined particle shape selection in isolated particle growth, with shapes defined by faces of the crystalline symmetry of the substrate, but with proportions different from those in the equilibrium Wulff growth. Such an assumption allows us to carry out numerical simulations for large enough particles and groups of particles to enable study of features of sintering. It also, however, represents a limitation because this idealization ignores the role of grain boundaries and their diffusion, which could be important in experimental situations especially for sintering of particles larger than nanosize.[30]

## 2.2. The Model and Its Numerical Implementation for Sintering

We consider the dynamics of nanocrystals of varying sizes with the simple-cubic (SC) lattice structure. Initially, before the onset of heating, their shapes were selected to correspond to equilibrium Wulff-construction configurations. As the temperature is raised, probability of atom detachments (evaporation) is increasing. The surrounding space then fills up with atoms which undergo continuous-space (off-lattice) Brownian motion. They can be recaptured at vacant lattice sites adjacent to the evolving particle surfaces, according to the following rule. As explained in Ref. 34, in order to suppress development of large-scale defects and emulate the tendency to maintain the particle crystalline structure, the atom attachment is "registered" with the SC lattice drawn continued to outside the particles. Nanocrystal morphologies of interest are in most cases obtained in the regime where large defects are dynamically avoided/dissolved, which is mimicked by our "exact registration" constraint. Each vacant lattice site which is a nearest-neighbor of at least one occupied site is surrounded by a conveniently defined "box" (we used the Wigner-Seitz unit-lattice cell). If an atom moves to a location within a box, it is captured and attached at the lattice location at the box's center. The model also incorporates the dynamics of the bound atoms, which can hop to neighboring vacant sites. The rates of such hopping events are determined by the number of the occupied neighboring sites at the initial and final positions (by the binding energy change).

The detailed kinetic rules are addressed shortly. Presently, let us outline our implementation of the off-lattice diffusion, the details of which are given elsewhere.[34,35] Diffusing atoms hop a distance equal one SC-lattice spacing $\ell$ per each time step, in a random direction. Hopping attempts into any SC cell which contains an occupied lattice site at its center, are failed. We use dimensionless units such that the time step of each kinetic MC sweep through the system and $\ell$ are set to 1. An atom that hops to any point inside a vacant cell which is a nearest neighbor of the nanoparticle structure, i.e., has at least one occupied SC-lattice cell nearby, attaches at the center of that vacant target cell.

Atoms which are parts of the nanoparticles' structure can hop to the nearby vacant lattice sites without losing contact with the main structure. Other hopping directions correspond to detachment, with the atom joining the freely-diffusing atom population. The set of the six possible displacement vectors for particle hopping, $\vec{e}_i$, included only those pointing to the nearest neighbors. However, the twelve next-nearest-neighbor displacements were considered in earlier modeling of isolated nanoparticle growth.[34] This choice was shown to have an effect on the nanocluster shape proportions.[34] The specific dynamical rules described shortly, follow those in the earlier work.[34] They mimic thermal-type transitions and are not corresponding to any actual physical interactions nor to any realistic kinetics. More realistic modeling would require prohibitive numerical resources and thus make it impractical to study large enough systems to observe the features of interest in neck development in sintering.

To define the kinetics, we note that each atom capable of hopping (means, an on-surface atom) will have at least one vacant neighbor site and thus the coordination number $m_0 = 1, 2, \ldots, 5$ (for nearest-neighbor SC). In each MC unit-time-step sweep through the system, in addition to attempting to move each free atom, we also attempt to hop each on-surface atom. The probability for a surface atom to move during a time step is $p_{mov} = p^{m_0}$. We assume a certain free-energy (per $kT$) barrier, $m_0 \delta > 0$, to overcome, and we expect that

$$p \sim e^{-\delta} < 1. \qquad (1)$$

Here $\delta \sim 1/kT$. However, the probability for the atom, if it moves, to hop to a specific vacant nearest-neighbor site will not be uniform but rather proportional to $e^{m_i |\varepsilon|/kT}$, where $\varepsilon < 0$ is a certain free-energy at the target site, the coordination of which, if selected and occupied, will be $m_i = 0, \ldots, 6$.

As common for simplified models of particle morphology evolution, our transition rules are not directly related to realistic atom-atom and atom-environment interactions or entropic effects, and, given that we are studying a nonequilibrium regime, no attempt has been made to ensure thermalization (to satisfy detailed balance, for instance). We expect that the surface diffusion rate is approximately proportional to $p$, is temperature-dependent, and reflects the effects of surface-binding energy barriers. The other parameter to vary in order to mimic the effects of changing the temperature, is

$$\alpha = |\varepsilon|/kT, \qquad (2)$$

which involves (free-)energy scales, $\varepsilon$, more related to the entropic properties. These expectations are primarily based on empirical observations. Studies of the growth of particles[34] with well-defined nanosize shapes for the SC lattice symmetry have yielded typical parameter values such as

$$p = 0.8, \ \alpha = 3.5. \qquad (3)$$

However, in the sintering situation with particles initially positioned with gaps of up to 6-7 lattice constants, $\ell$, and without the external supply of atoms, formation of necks between adjacent particles is improbable for these parameter values, because the equilibrium concentration of atoms maintained by evaporation off the particles in found small. To initiate neck formation, the system had to be heated to a sufficiently high temperature, $T'$. As described below, numerically we found that the corresponding lower values are $\alpha' = 1.8 - 2.4$, with the respectively adjustment of $p' = p^{\alpha'/\alpha}$, see the discussion above in relation to Eqs. (1)-(2).

Simulation of the dynamics of sintering requires consideration of nanoparticles which each contain from a fraction of a million to a couple of millions of atoms. Therefore, we had to limit the present work to consideration of local configurations with initially up to 5 nanoparticles. The linear span of the particles did not exceed 100 (in our dimensionless units, per lattice spacing $\ell$). In order to model the fact that the simulated representative local particle configuration is actually a part of the larger system, we used the following approximation. Each of the considered particles, the shape of which initially has Wulff-construction faces, was enclosed in a cube bound by (100) type faces. These faces were defined at distances of ~ 8-10 lattice spacings from the (100)-type face tangential to the particle. Atoms which evaporated off the considered particles were not allowed to leave the space defined by the set of these cubes for our nanoparticle cluster. Except for the described modifications, the model was otherwise the same as detailed in earlier works.[34-36]

Our earlier presentation emphasized the fact that the presently adopted kinetic MC approach utilizes dynamical rules only suitable for isomeric *nanosize* particle shapes, in order to make the simulations tractable for particles sizes of relevance. Surface evolution at the nanoscale, for surfaces of linear sizes of several tens of atoms, can dynamically yield structures[34-36] different from those found for larger surfaces of linear sizes of several hundred of atoms and larger. The larger surfaces will typically develop ziggurat and/or step-patterned structures during their evolution/growth driven by thermal-type fluctuations.[51] We find that such morphology is beneficial for faster neck formation and is present for smaller surfaces as well. However, for some configurations of smaller surfaces facing each other, layering is the dominant mode, which delays neck formation. Two possible initial configurations of a group of particles to be used to study neck formation are illustrated in Figure 1. Since our "atoms" are modeled as point-like objects, sizes/shapes of the symbols depicting atom/particle configurations in all the figures are for illustration purposes only, to allow detailed local structure inspection.

An ideal (100)-type face increases the binding energy by $3|\varepsilon|$ per atom as compared to a dilute solution (in which the separate atoms would also acquire a free-energy contribution due to entropy of mixing). For possible types of (110) and (111) surface planes the same answer, $3|\varepsilon|$, is obtained.[34] This energy is simply the binding energy per atom in the interior of the SC structure. However, the energy of evaporation (the average number of the locally broken bonds) per atom, $\varepsilon_{ev}$, from the fully packed faces is different: $\varepsilon_{ev} = 5|\varepsilon|$ for (100), $\varepsilon_{ev} = 4|\varepsilon|$ for (110), and $\varepsilon_{ev} = 3|\varepsilon|$ for (111). Similarly, the average energy of atom attachment on top of a

fully packed layer, $\varepsilon_{ad}$, is also different: $\varepsilon_{ad} = |\varepsilon|$ for (100), $\varepsilon_{ad} = 2|\varepsilon|$ for (110), and $\varepsilon_{ad} = 3|\varepsilon|$ for (111). Atom evaporation rate will on average be in the order (111) > (110) > (100). As the atom concentration in the surrounding medium increases, the supersaturation will first be reached for the (100) faces. However, atoms attaching on the (100) type faces are less bound. The hopping probabilities, $p_{mov}$, and therefore the surface diffusion rates, will be in the order (100) > (110) > (111). For an isolated nanosize particle, fixed shapes can approximately persist over periods of time because flux of matter to the (100) faces, for instance, via the surrounding medium can be compensated by the on-surface diffusional transport. For particles in close proximity to each other, exchange of matter can modify this picture, leading to neck formation, sintering and other effects.

## 3. Results and Discussion

### 3.1. Layering Mechanism for Neck Formation at the Nanoscale

Dynamics of particles facing each other with (100)-type surfaces can be considered on an example shown in Figure 1(a). We will argue for an initiation mechanism of neck formation by forming new layers at these facing surfaces. The particle "radii," defined by half-spans along the (100)-type directions, were $R=50$ for each of the two outer particles (A,C), and $r=30$ for the middle particle (B). Our choices of these and other parameters are discussed below, and are aimed at obtaining sintering configurations likely to be of interest as "cartoon models" of experimental situations (where useful parameter values are presently found largely by trial and error). Note, for example, that for $r \leq 25$ (with all the other parameters, given shortly, fixed) no merging of the smaller particle with the larger ones could be obtained in our numerical experiments.

We note that sintering of specifically noble metal nanoparticle of recent interest in improved conductive layer preparation, involves complicated steps starting with particle synthesis which usually leaves organic residues at their surfaces (such as Arabic gum), and then suspending of particles in a viscous solvent (paste) which typically also contains other fillers and is printed as "ink" on a substrate, as well as shaking/tapping/compression at various stages before the actual sintering, typically done together with "firing" which burns away some of the organics.[23-26] Experimental data on such a pre-sintering (pre-firing) "green" (tapped/compressed) density for nanoparticles of the considered nanosizes, suggests that an assumption of interparticle gaps of up to order 3 to 5 atomic layers is more realistic than direct contact (specifically due to the presence of stabilizing organics left over from synthesis).[23] Furthermore, many of the fillers remain after the sintering process.[4,23-26] The "contact formation" step is also distinctly identified in some multiscale-approach studies to sintering.[13] We took both gaps here as initially $\Delta = 4$ unfilled lattice planes, i.e., the physical distance of $5\ell$, and the dynamical parameters at elevated temperature, defined in connection with Eqs. (2)-(3), were $\alpha' = 2.2$, $p' = 0.87$.

Figure 2 shows a fluctuation event of a link forming through layers **a**, **b**, **c**, **d**, connecting planes **1** and **1'**, and thus bridging particles A and B in the configuration of Figure 1(a). The inset in Figure 2 shows a large-time configuration of a MC run for which well-defined necks were already formed connecting all the three particles. However, the initial bridging fluctuations occur for much shorter times, by forming and then dissolving needle-shaped bridging links, to be termed "needles" for brevity. The first needle spanning layers **a**, **b**, **c**, **d**, formed at time $t = 28 \times 10^3$, see Figure 2. This needle rapidly evaporated and had little probability of persisting for extended times. A needle can evolve into a stable neck (one surviving and coarsening for prolonged times of the numerical simulation) only if it is rather short, means, only after the gap between the particles was narrowed. The preliminary step of narrowing of the gap between the (100)-type faces occurs by the formation of additional densely packed layers. We found that layer **a** (on the larger particle) is more likely to fill out than layer **d** (on the smaller particle).

This observation is reminiscent of the phenomenon of Ostwald ripening.[52] Larger particles are more efficient in the consumption of the diffusing atoms than smaller particles. Here the mechanism is more local. Indeed, as already mentioned the saturated-diffuser concentration is somewhat different for different-symmetry crystal faces. Once the temperature is elevated, matter evaporating from the (110) and (111)-type faces will be transported to replenish the supersaturated diffuser "gas" in the gap between the (100) faces of the particles. Because the gap is small, these diffusing atoms are likely to be captured at the growing surfaces emerging at layers **1** and **1'**. There is also local detachment into the gap, leading to the direct exchange of matter between these surfaces. On-surface cluster formation, etc., can occur, as studied for the FCC symmetry in Refs. 35, 36.

In Figure 3, we show that in the course of a typical MC run for short times, layer **1** of particle A had many vacancies formed in it. Most of these, highlighted as magenta circles, were later filled by atoms transported form particle B. A cluster of atoms emerging on top of this layer (olive spheres) is initially unstable, see the sequence of Figures 3(a)-(c). As long as transport processes favor growth of these faces, **1** and also **1'**, ultimately clusters on them exceed a certain critical size and become stable. This has been reported in a study of an on-surface growth in Refs. 35, 36. The onset of cluster growth and ultimately merger to create new layers, can be seen in Figure 2 (the olive curve for layer **a**). It sets in starting at $t = 35 \times 10^3$, presumably with the cluster shown in Figure 3(c).

However, there is an important difference here as compared to the on-surface growth considered (for FCC rather than SC) in Refs. 32, 33. In that study, clusters forming as islands were overgrown as pyramid-shaped formations before and during them merging to complete new layers. Results such as those shown in Figure 2, suggest that here the in-layer cluster grows while remaining primarily two-dimensional, in layer **a**, with little matter present in layers **b**, …, **d**. Instead, loss of matter at the periphery of layer **1** causes a noticeable decrease in the total number of particles in it (see the magenta curve in Figure 2). The diffusional flux that would replenish this layer to yield the steady-state shape for an isolated particle, is now redirected to the building up of layer **a**. This can be seen for times $4 \times 10^4 < t < 5 \times 10^4$. Further interesting processes occur for times $5 \times 10^4 < t < 1.2 \times 10^5$. During this time interval, layer **a** completes its formation, but at the same time layer **1** is somewhat replenished at the expense of layers behind it. This represents the onset of a neck-like distortion of the initially flat (100) cluster face. At this point the gap between the particles is approximately three layers, and needles develop which span the gap for short times, as described earlier.

We note that the processes in the outer largely empty layers, here **b** and **d**, become correlated when they are very close. In this example, these two layers fill up at the same time, starting at $t = 1.2 \times 10^5$. However, this is accompanied by the formation of layer **c** (not shown in Figure 2) which completes the neck. This is illustrated in Figure 4, which shows the first bridging (by a single atom in layer **c**) at $t = 1.23 \times 10^5$ leading to the formation of the stable neck. At this stage, layer **a**, for instance, consists of approximately 25% of atoms originating form particle B and none from C, confirming that flow of matter in neck formation is primarily local. All these observations are actually realization-dependent, because the described growth phenomena are fluctuation-driven. The time scales and the order in which surface layers fill up at the two particle surfaces can vary between MC runs. However, the general observation mentioned earlier has been that on average the larger of the already formed layers facing each other are more likely to have the next layer emerge on top of them first.

Once formed, the stable neck grows very fast, reminiscent of supercritical cluster growth in nucleation. Figure 5 shows the neck structure soon after it is formed and the time dependence of the number of atoms in various layers adjacent to the neck. As noted earlier, most of the depletion of the matter from the underlying layers to form the neck occurs at the larger of the two facing surfaces. The formed neck cross-sectional shape has a square shape corresponding[34] to the nonequilibrium growth for the SC lattice symmetry. The equilibrium shape would tend to have an octagonal cross-section.

Figure 6 illustrates the property that exchange of matter, proceeding by atom evaporation into the "gas" of diffusers, leads to the incorporation of the "foreign" atoms not only at the surfaces which evolved from the original particles' faces, but also in the interior of their structures. Indeed, there is a substantial exchange of matter between the original particles. For instance, in Figure 6, the percentage of atoms originating from particle B at $x = 134$, which is deep in its original structure, is only 83%. The remaining 17% came from particles A and C. As the system is heated, the crystal structure, especially at the surfaces, develops local, small-size fluctuations, pits, vacancies, cavities, etc. The forming vacancies then diffuse and spread into the interior of the structure. This facilitates atom mixing inside the structure in addition to surface dynamics. We observed that many of the incorporated atoms have coordination numbers less than 6 and are therefore immediately movable. This is particularly true at surfaces, and as a result the cross-sections of the layers interior to particle A on its side from which the neck emerges, for instance, are significantly distorted from their initial nearly square octagonal shapes and are nearly circles, which corresponds to quasi-equilibrium for high temperatures. This is true even for layers that are quite far from the neck, e.g., at $x \sim 75$, in the notation of Figure 6.

Further increase in the temperature leads to acceleration of all the dynamical processes in the system. However, generally the mechanism of neck formation remains the same. The base layers, **1** and **1'**, remain relatively densely packed, see Figure 7. However, the increase in the temperature has led to a larger flux of atoms into the gap. This significantly shortens the formation of the stable neck, from $t = 1.2 \times 10^5$ for $\alpha' = 2.2$, to $2.7 \times 10^3$ for $\alpha' = 1.8$. In fact, fast exchange of atoms ongoing between particles A and B leads to a significant correlation in the formation of smaller clusters on top of the particle surfaces: Compare the red and blue clusters in Figure 7 at various times. At $t = 2.7 \times 10^3$ two contacts develop between A and B, though at a later time the left contact and the accompanying clusters dissolve. It is important to note that this behavior is a nanoparticle feature. For larger particle faces, we expect that elevation of temperature would "roughen" their surfaces and eventually all of them would evolve into multi-cluster fluctuating ziggurat and/or step-patterned structures.[51]

Furthermore, for the layering-mechanism, as opposed to the self-clustering mechanism considered in the next section, neck development for sintering of two particles was found to depend markedly on their sizes. For example, if the size $r$ of particle B is decreased, the gap area of exchange of atoms between the facing layers **1** and **1'** shrinks, and as a result, the lifetime of atoms adsorbed on top of layer **1'** is shortened. They can roll off it, diffusing to the nearby (110) and (111)-type faces, on time scales $\sim r^2$. Specifically, for $\alpha' = 2.2$ and $R = 50$, sintering would not actually occur for the system geometry shown in Figure 1(a), had we have taken $r \leq 25$.

Recall that we took $r=30$ for Figures 2-6. To get sintering in the former case, temperature has to be increased.

### 3.2. Cluster-Growth Mechanism for Neck Formation

Let us consider the case of the particles pairwise facing each other with (110)-type planes; see Figure 1(b). The length of the gap is still assumed $7\ell/\sqrt{2} \sim 5\ell$, though in this orientation six (110)-type lattice planes are contained in the vacant space. These are marked **a**, **b**, **c**, **d**, **e**, **f**, in the figure, counting from particle A to B. Similarly, the filled layers, such as **1**, **2**, … and **1'**, **2'**, … are counted/labeled as the (110)-type lattice planes rather than physical distances in terms of $\ell$.

Figure 8 provides an account on the processes ongoing near the surface of particle A, cf. Figure 1(b), facing the gap. As the number of atoms in layer **1** decreases, these atoms are consumed to form few-layer clusters in **1**, **a**, and **b**, initially mostly extending to layer **a**, then **b**, and layer **c**. This is similar to the standard surface growth.[51] Here we offer an explanation of why such a "layered pyramidal cluster structure" emerges at sintering (110)-type faces but not for (100)-type faces described in the preceding subsection.

In Subsection 2.2, we noted that the energy of evaporation out of a densely packed (110) face, which is stepped, is $\varepsilon_{ev} = 4|\varepsilon|$, lower than $5|\varepsilon|$ for (100). However, the energy of attachment on top a densely packed (110) face, is $\varepsilon_{ad} = 2|\varepsilon|$, larger than $|\varepsilon|$ for (100). These energy differences apparently cause the sintering scenario of nanosize faces to become cluster-type growth for (110) for the same temperature for which it is layer-type for (100). Figure 9 shows the morphology of the linking clusters for three times during the process. In the (110)-type sintering configuration of Figure 1(b), the flux of atoms out of the originally filled layers, notably, **1** and **1'**, is much more profound than in the (100) case, and the layer illustrated in the figure is no longer dense: It became a part of the protruding/connecting cluster structure. We also note that (110)-type faces are less isotropic than the (100)-type faces, and as a result the clusters are somewhat elongated in the direction along which the original (110) face is stepped; see Figure 9.

Figure 10 gives interesting statistics for cluster-driven sintering. The process is primarily driven by on-surface diffusion, including that from faces nearest to the gap. Supply of matter into the neck-formation region from the outside, via the "gas" of diffusing atoms originating from other particle faces is limited on the fast link-formation time scales. Figure 10 illustrates that the initial evaporation is obviously followed by matter being transported towards the gap, starting approximately on the time scales, $\sim 10^3$, of the onset of the formation of the linking clusters (marked in the figure). We also observe that after a time of approximately $3 \times 10^3$, the faction of atoms in the neck-formation region which never detached, stabilizes at $\sim 60\%$ (the evaporation affected the dynamics of only $\sim 40\%$ of the atoms), whereas the role of evaporation is still increasing at the other (110)-type faces.

The most important change in the sintering process due to this cluster-formation driven neck development has been that at the same temperature, the sintering time here is much faster than for the layer-formation driven process. The time scales of the formation of a substantial neck are reduced by 2-3 orders of magnitude. Given such a fast emergence of well-developed necks for certain configurations of particles facing each other, one can question what other dynamical changes might occur, and will these formed necks remain intact, on much larger time scales, if the system is being heated up for longer times in an attempt to sinter the other gaps between various particles. This is considered in the next subsection.

### 3.3. Dynamics and Stability of Contacts Mediated by Small Particles

At larger time of evolution, necks formed by mechanisms described in the preceding sections can break, and furthermore, global changes in the system's structure can occur. This can be particularly important for situations when particle sizes are not comparable. Specifically, the recent experimental finding[4] that small particles of radii such that they snuggly fit in the voids between larger particles, can mediate better connectivity/neck formation, should be examined in this context. Indeed, especially for nanoparticles, we have already observed that time scales of neck development may differ by orders of magnitude depending on the local configuration. Furthermore, we also found that Ostwald-ripening type coarsening processes initiated at the facing surfaces occur on the time scales comparable to those of the slower neck formation. Therefore, the structure the smaller particles may significantly change, and they might not even survive as independent entities during the sintering process. This effect is illustrated in Figure 11, which is discussed in detail shortly.

For small particles placed in the voids between larger particles, ideally, we would want these necks to ultimately cause the smaller particles to distort into cylindrical bridges connecting the large particles. The desirable bridging structure might be more complicated in more realistic geometries, if a small particle is positioned between more than two larger particles. For example, for spherical particles in random close packing, the voids typically accommodate small spheres touching four large spheres. As mentioned, the formed bridges mediate flow of matter between the large particles and therefore can "fatten," largely driven by surface diffusion. However, evaporation resulting in exchange of matter via the diffuser "gas" of atoms, can dissolve them. Note that for the same radius, spheres have twice larger mean curvature than cylinders, i.e., they are less stable with respect to evaporation. Therefore, elongated structures forming as connectors between isometric particles can at least locally become stable against evaporation if they are "fattened" fast enough by the supply of matter from the large particles. This will occur for cross-sectional sizes smaller than those of the particles that they connect. Dumbbell structures might therefore be possible to obtain for carefully designed sintering processes. In most cases, the bridging particles are actually quite small. For example, based on random close packing type considerations, experimental work[4] reported the use of particles of radii $r \sim R/7$. These are unlikely to contain enough matter for the stabilization effect upon elongation into bridging regions. Therefore, presumably they can provide improved connectivity, as reported,[4] only under conditions facilitating fast flow of matter between the particles being sintered rather than via the surrounding medium.

Figure 11 illustrates the dynamics of a neck formed and later dissolved, in the geometry of fast neck formation, cf. Figure 1(b). Here we took a smaller value, $r=20$, than in Subsection 3.2, but the time scale of the initial neck formation via the clustering mechanism remained nearly the same. After the initial neck formation, the contact regions A–B and B–C, the cross-sections of which are displayed in the figure, are fattened, and the whole connecting shape becomes more and more cylindrical. The data presented in Figure 11, suggests that as long as they are narrow and therefore have large negative curvature, the dynamics of these regions is primarily at the expense of atoms from particle B, presumably coming from its less bound surface layers. At a later stage, mixing of atoms form the large and small particles occurs in the connecting region, evidencing a significant transport of matter all across the connecting structure.

It transpires that surface diffusion is the main driving mechanism decisive for the survival of the connecting structure at later times, cf. Figures 11(c)-(d). Fluxes of matter via this mechanism are directed towards the regions of the highest negative curvature along the connecting structure: yellow arrows, $\Gamma^-$, and white arrows, $\Gamma^+$, see Figure 11(c). By simulating the system with various parameter selections, not reproduced here, we observed that there are two competing trends in the connecting region evolution. Irrespective of the relative magnitude of the two fluxes, the two high-negative-curvature regions, initiating from the original gaps, not only fill up and smooth out but also move towards each other. The former process: filling up/smoothing out, is controlled by the sum of the two fluxes. The latter process of moving towards each other, is controlled by the difference, $\Gamma^+ - \Gamma^-$.

This evolution is accompanied by the general shrinkage of the part of the whole central structure that is still identifiable with the formerly small particle (B). At the same time, the trend sets in that, while initially, $\Gamma^- > \Gamma^+$, the flux $\Gamma^-$ weakens and can become significantly smaller than $\Gamma^+$ for later times, as the cross-section becomes more cylindrical. The interplay of these time-dependent changes in the system structure and transport in it, determines whether the connecting region survives or not. Figure 11 illustrates the latter situation whereby the $\Gamma^-$ flux did not weaken early enough and therefore the connection ultimately broke. Obviously, this "cartoon" description is at best approximate and applies for a limited time interval of the existence of all the mentioned structures: the central region still identifiable with what was particle B, and the negatively curved connecting regions.

The disappearance of the contact can be avoided by affecting the balance of the two surface fluxes. One way to accomplish this, is by increasing the temperature, corresponding to decreasing $α'$, as illustrated in Figure 12(a). Apparently, changes in the mobility of atoms at the various types of surface regions involved, cause the connecting region to stabilize, rather than break at time $t_{break}$. For the parameter values of Figure 12(a), this apparent "dynamical phase transition" occurs at approximately $α' = 2.2$. Another way to affect the balance of the fluxes, is by changing particles sizes. Figure 13 illustrates an interesting property that when the smaller particle (B) size was reduced as compared to that used in Figure 12(a), with all the other parameters the same, then no stable connection survived for increased temperatures and for large times studied. However, when the larger particles (A and C) were also reduced in size, the transition to a stable connection for high enough temperatures was found again, see Figure 12(b).

Figure 14 illustrates the dynamics of a more complicated particle arrangement. Here the four larger particles positioned in a square arrangement with gaps of $12\ell$, were found to never form any connections on their own for the selected system parameters and time scales of the simulations. However, an added small particle in the central void, see Figure 14, facilitates the formation of a single sintered entity on the time scales comparable to those found for the same initial gaps and facing crystalline surfaces for the configuration of Figure 1(b). The added smaller particle serves to initiate the link formation with all the surrounding larger particles. The formed connections ultimately merge into a single negative-curvature central region for the sintered entity on time scales one order of magnitude larger. For very large times the structure is expected to further evolve as single more isometric entity, but this was not observed in the largest of our simulation runs.

By changing the particle arrangement and gaps between them, one can create various scenarios for the role of the small particles as mediators of the formation of connection structures between large particles. We will consider a couple of illustrative examples. We note that ideally in sintering, one would want to place the large particles initially in as dense a configuration as possible. However, then the smaller particles fitting in the remaining voids will have fractionally smaller average diameters on the scale of the larger particle sizes. Figure 15 provides an interesting illustration of a configuration somewhat denser than that considered before. The configuration is asymmetrical, and without the small particle only the neck bridging the two closer-positioned large particles (see Figure 15) would ever form. Adding the small particle affected the overall necking configuration and facilitated formation of another neck, making the whole structure connected. However, because the added particle is now relatively smaller, in order to ensure stable connection formation rather than this particle dissolving, higher temperature was used, corresponding to $\alpha' = 1.5$, as a trade-off.

As the added particles become relatively smaller in denser configurations, the random statistical nature of the sintering process, driven by thermal fluctuations, becomes more apparent. This is illustrated in Figure 16. Here the initial configuration is symmetrical, but random fluctuations "break the symmetry" for later times, resulting in this case in an asymmetrical configuration of the evolving necks between two of the large particles, formed mediated by the small particle. However, such a stable neck formation was not as probable as might be desirable in applications. Indeed, for approximately 70% of the random realizations, with the same initial configuration shown in Figure 16(a), the small particle was simply dissolved and merged into one of the two left particles.

Variation of the temperature can also affect the patterns of the most probable evolution and the resulting possible neck configurations. This is illustrated in Figure 17. For various temperatures, the small particle can be dissolved into one of the larger ones: see panel (a). For other temperatures, neck formation can be initiated and then pairwise bridge one of the two left and the right particles: panel (b). Or, the two left particles can be bridged: panel (c).

## 4. Conclusion

Sintering is a complicated process modeling of which requires a multi-scale approach. Indeed, local bridging at the (near-)contacting particle surfaces results in the initial emergence of

– 20 –

necks. We observed that these can form on vastly varying time scales depending on the local configuration of the facing particle surfaces. The necks then coarsen, while the connected structure still retains the local identity of at least the larger original particles. A later stage is expected when the whole structure further compactifies, and even the larger particles lose their identity as separate entities. New effects are expected at this final sintering stage which were not observed in our simulations, because, in order to make them numerically tractable, we had to limit them to local geometries of a couple of particles only, and to MC runs not long enough to observe the large-time particle merging.

One experimentally well-known property is that the connected larger samples or layers being sintered, at the later stages will noticeably compactify, typically losing up to order 25% of their volume. To check the extent to which our approach is at least potentially realistic for such late stages, we ran a couple of particularly large (in terms of the numerical resources required) simulations of just two larger particles (size 100, gap 4, $\alpha' = 1.8$), without any added smaller particle(s), sintering to the late stage when they actually merged into a connected "peanut shaped" dumbbell structure with the neck nearly as large (in transverse size) as the dimensions of the two end masses that originated form the initial particles. It was found that at this stage, all the linear dimensions of the connected structure (transverse and longitudinal) shrunk by about 10% as compared to the original configuration (of two close but disconnected particles), suggesting that indeed the model can reproduce order 30% volume shrinkage.

Furthermore, at this stage of the dynamics, only approximately 1% of the atoms where in the diffuser "gas" and therefore their evaporation could not be blamed for the loss of the connected structure size, which is therefore largely attributable to the geometrical restructuring associated with the particles' merger. However, as noted earlier, our idealization of not accounting for the role of grain boundaries which can act as the sources and sinks of vacancies and thus affect densification, and, in fact, can also alter the structure and stability of the bridging regions, suggests that a direct comparison with experimental numbers for volume shrinkage might require a more detailed and presently less numerically tractable modeling approach. In addition, particle movement with respect to each other and effective "forces" between particles as driving their dynamics,[10,53] could then possibly also be considered.

In summary, the developed numerical approach allows modeling of certain features of the short-term, local geometrical processes occurring during sintering at the nanoscale. Specifically, we considered the aspects of sintering of nanosize particles related to the competition of several mechanisms of direct local bridging or neck formation mediated by an added smaller particle. One of the interesting new findings has been that for certain types of facing surfaces the traditional growth mechanism via "cluster formation," which for larger planar surfaces would be a ziggurat/step-patterned growth, dominates. For some other surfaces, those with insufficiently large difference between certain atom detachment and adsorption free-energies, layer-formation growth becomes dominant. The latter process is likely dominant only at the nanoscale and requires significantly larger temperatures/longer times for sintering than the clustering process. The main practical implication of this finding is that for nanoparticles which are synthesized highly crystalline,[54] with well-controlled rather than randomly present crystalline faces, the shapes and face-symmetry selections can play an important role in determining the ease of getting them sintered.

Another finding has been that particles' faces' proximity has a dramatic effect on their sintering. The initial gaps should really be at most few crystal spacings wide in order for bridging to develop on reasonable time scales. Establishing such initially very small gaps can be mediated by adding smaller particles. However, there are some trade-offs. Specifically, the smaller particles, when added as part of the mixture, can prevent the larger particles from packing more densely. If they are made too small to avoid the latter effect, the added particles can dissolve, with their matter incorporated into the coarsening larger particles, rather than mediate neck formation. It is important to emphasize that while the present approach illuminates certain aspects of nanoparticle sintering in situations when initial configurations are of the considered type, of relevance to specific experimental community,[4,23-26] it is limited, as are all the particular-size-scale modeling techniques. Our approach will hopefully add a new methodology within a general multi-scale theoretical understanding of the complexities of sintering.

At the nanoscale, fluctuations are significant enough to lead to possible different configurations and outcomes for the same initial particle arrangements, especially for processes mediated by the presence of the smaller particles. Generally, the presence of the added small particles is therefore in itself not a guarantee for improving the sintering process, such as allowing the use of lower temperatures or making the process faster. Our findings based on the shown configurations in various figures in this article, as well as many other numerical simulation runs, suggest that the key process step to control, is the initial effective creation of a sufficient number of well-established necks throughout the system. Local neck formation can require different temperatures and times depending on the specific configuration present at the interparticle near-contact gaps. However, once the first batch of necks was formed, coarsening of such necks into substantial bridges between the particles was found to be a more uniform process as far as temperature and process-duration effects are concerned. The best approach to a trial-and-error optimization of sintering of nanoparticle mixtures should be focused on devising the mixture composition and the initial heating protocol to form a connected structure. The later coarsening stages (mentioned at the start of this section) leading to a better product connectivity are then not too sensitive to this initial preparation step and can be adjusted/optimized separately.

**Acknowledgements**

The authors thank Prof. D. V. Goia, Dr. I. Halaciuga, Dr. I. Sevonkaev, and Dr. O. Zavalov for rewarding scientific interactions and collaboration, and acknowledge research support by the GISELA project, funded by the European Commission under the Grant Agreement no. 261487, and funding by the US ARO under Grant W911NF-05-1-0339.

53. K. Saitoh, A. Bodrova, H. Hayakawa and N. V. Brilliantov, *Phys. Rev. Lett.*, 2010, **105**, 238001.

54. I. Sevonkaev, V. Privman and D. Goia, *J. Solid State Electrochem.*, 2013, **17**, 279.
– 28 –

**FIGURES and CAPTIONS**

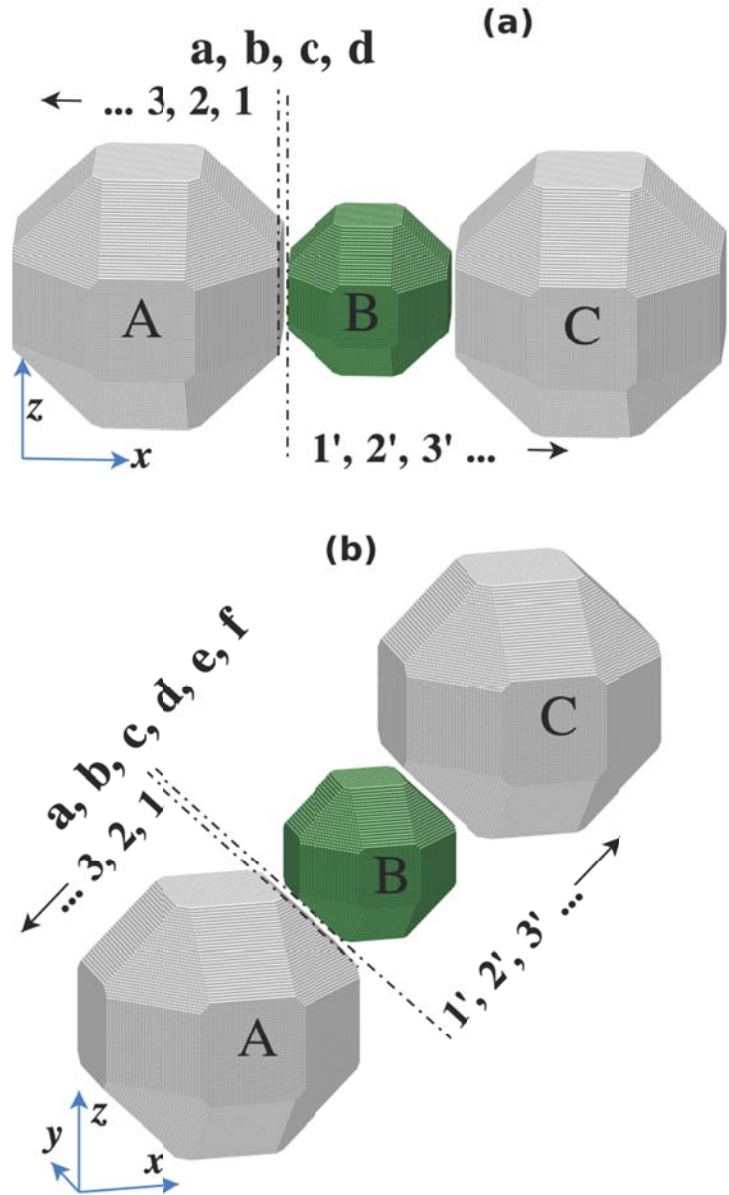

**Figure 1.** Possible initial placements of two larger particles (A, C) with the third, smaller one (B) in between, facing each other with the (a) (100)-type, and (b) (110)-type crystalline faces. The shapes are of the Wulff construction. The gaps correspond to several vacant lattice planes, marked **a**, **b**, **c**, **d**, …. The lattice layers adjacent to the gaps within the particles on the left of it, are labeled **1**, **2**, **3**, **4**, …, counting away from the gap, and to the right of the gap **1'**, **2'**, **3'**, **4'**, ….

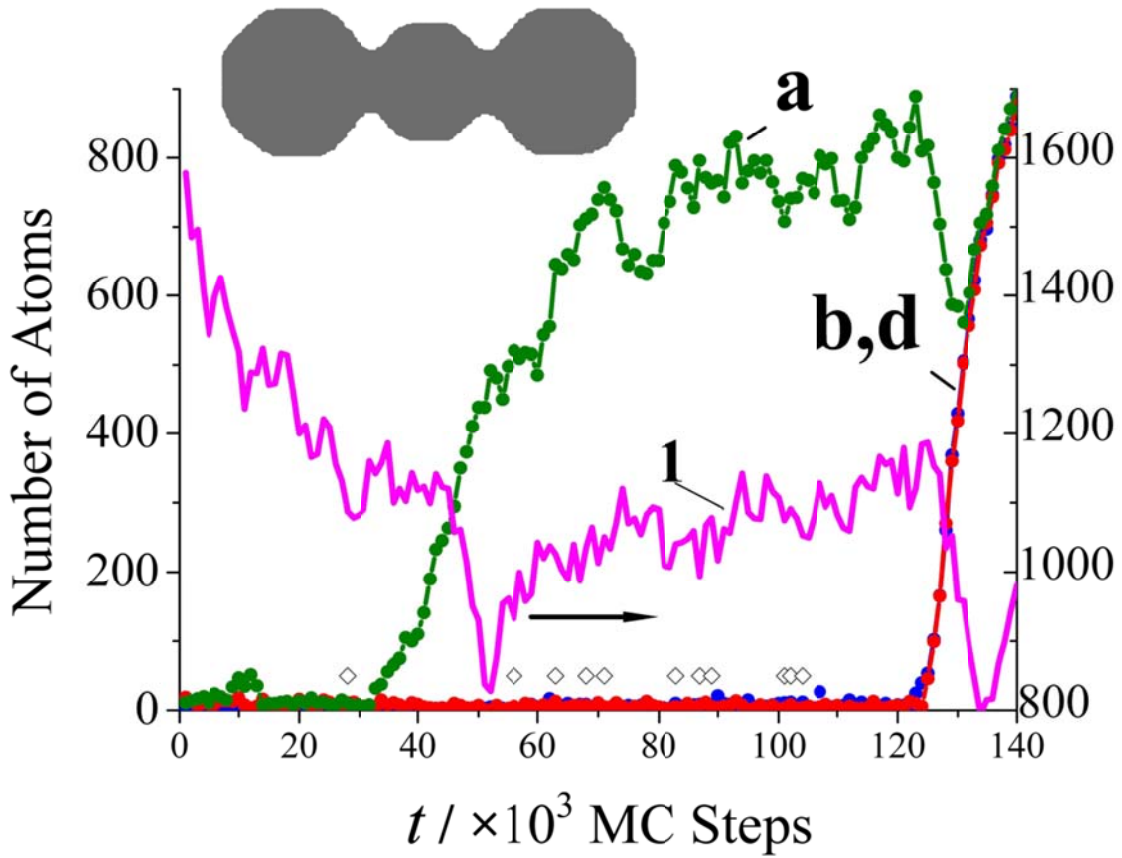

**Figure 2.** Time-dependence of the number of atoms in layer **1**, cf. Figure 1(a), for a typical MC run: shown in magenta, with the vertical axis on the right. As this and the nearby underlying layers lose a fraction of their atoms, the gap-layers begin to fill up: The left vertical axis shows the number of atoms in the initially vacant layers **a** (olive), **b** (blue), and **d** (red), respectively. Empty diamond shapes above the horizontal time axis mark instances of time at which particles A and B were temporarily bridged by a needle-shaped link, see text for details. Inset: cross-section of the system configuration after a large time of $t = 2 \times 10^5$.

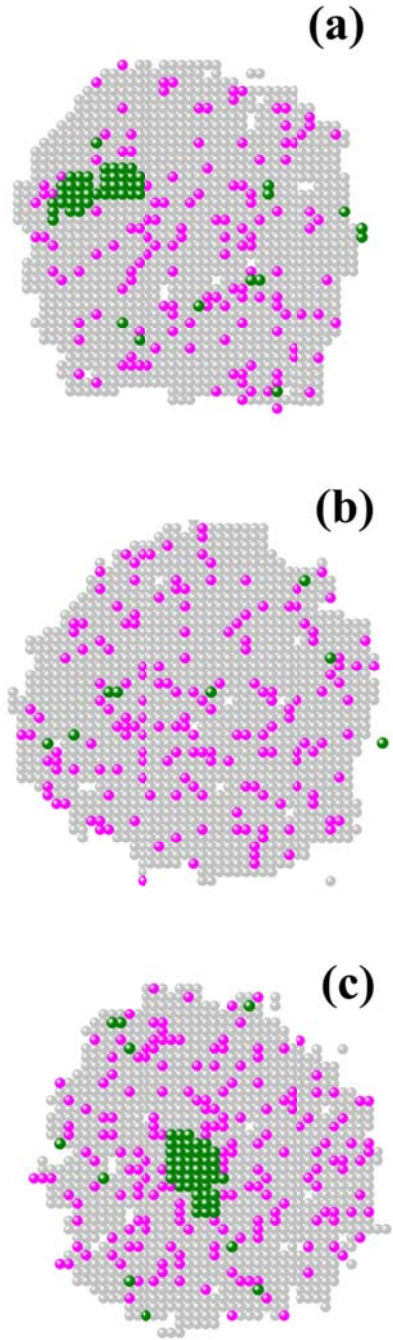

**Figure 3**. Example of evolution of the near-contact surface of particle A, in the configuration shown in Figure 1(a). Atoms in the originally outer layer, **1**, are shown in light gray and magenta. Those highlighted as magenta were initially in particle B but are now in layer **1** of particle A. The olive spheres show all the atoms in layer **a**, irrespective of their origin. The panels correspond to simulations times (a) $t = 12 \times 10^3$, (b) $14 \times 10^3$, and (c) $35 \times 10^3$.

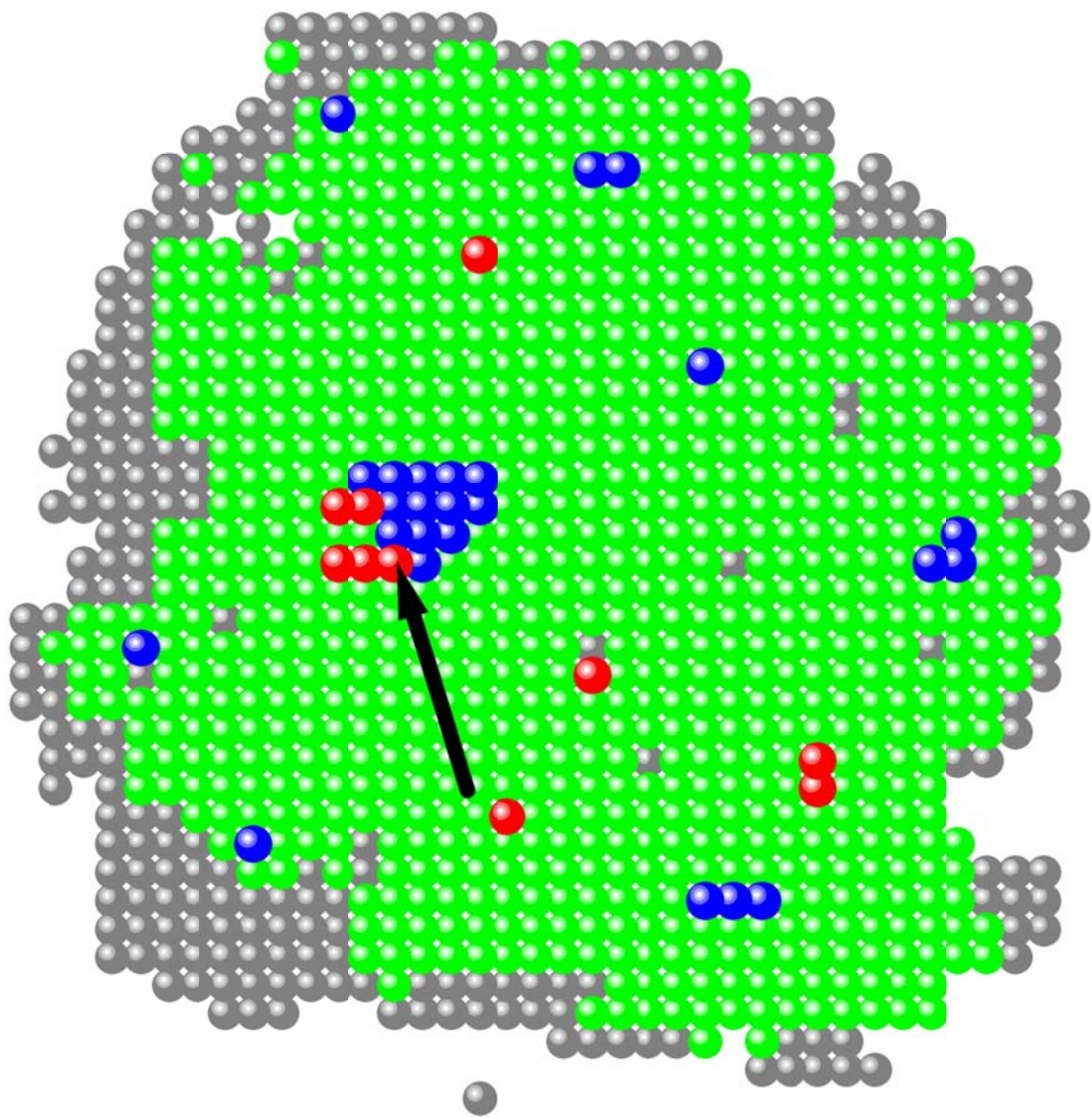

**Figure 4**. Atoms in various layers at the instance of the formation, $t = 1.23 \times 10^5$, of the stable single-atom bridge which later develops into a stable connection (neck) between particles A and B, in the configuration shown in Figure 1(a). Gray: layer **1**; green: layer **a**; blue: layer **b**; red: layer **d**. Layers **b** and **d** are connected by a single atom in layer **c**, which is under the red sphere marked by the arrow.

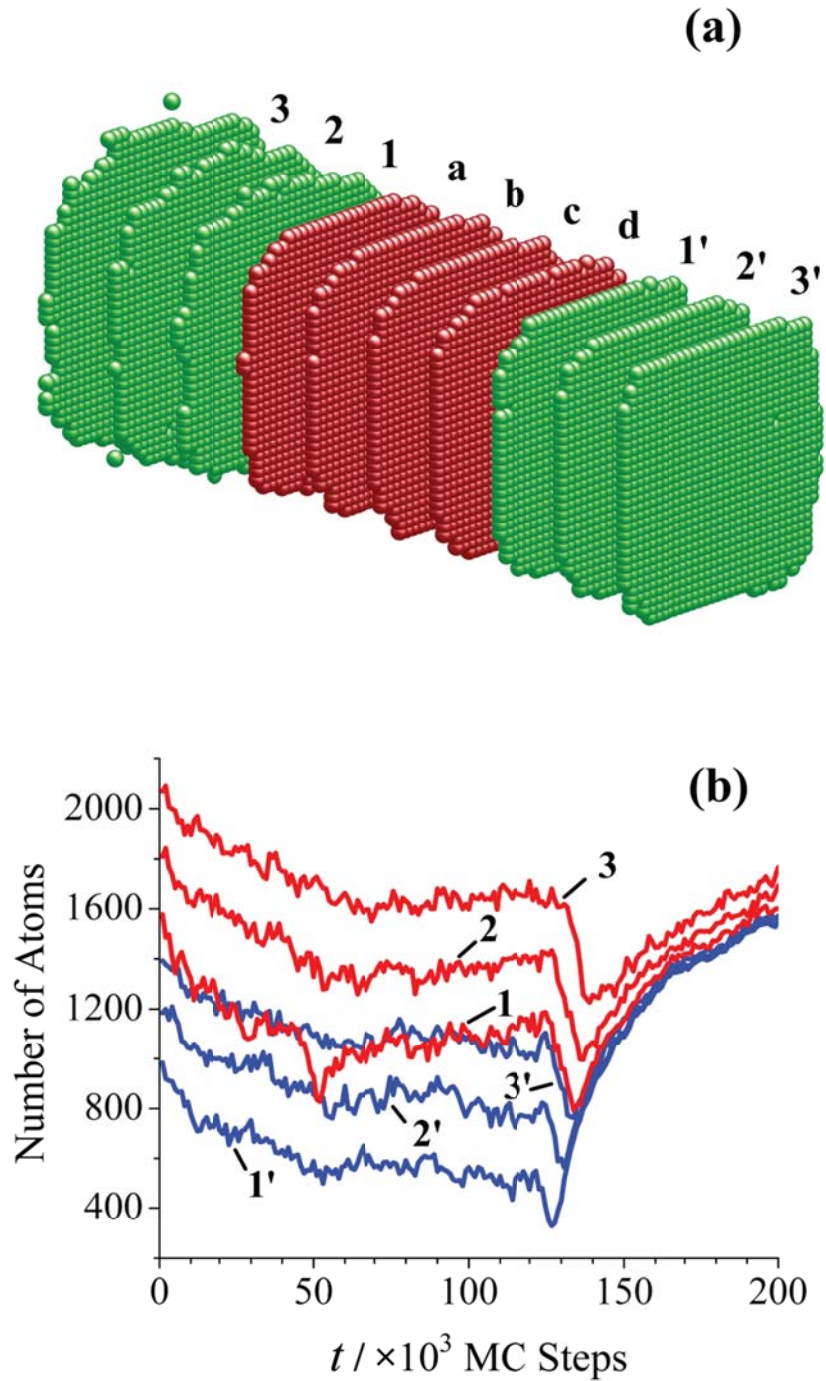

**Figure 5**. (a) The structure of the neck at time step $1.40 \times 10^5$, and (b) the time dependence of the number of atoms in the three neck-facing layers of particles A and B, in the configuration shown in Figure 1(a).

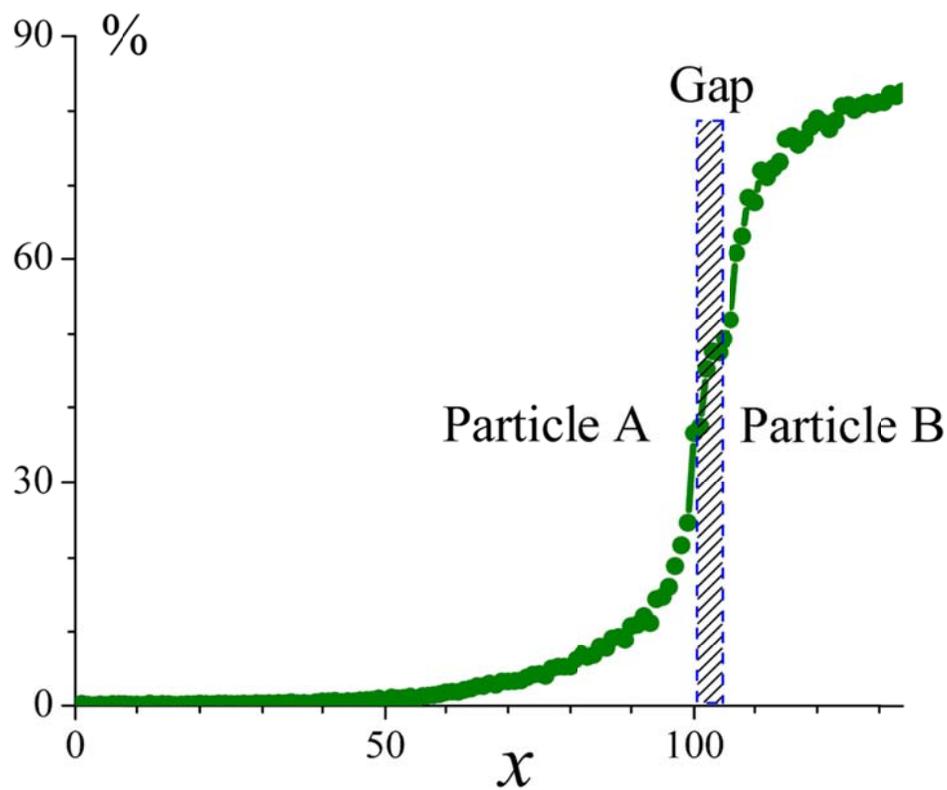

**Figure 6**. Illustration of the penetration of atoms from particle B into the structure of particle A at time $2.00 \times 10^5$. Here $x$ measures the displacement form the initial position of the left side of particle A, in the configuration shown in Figure 1(a), continuing into the gap (delineated by the hatched bar), and then into particle B. The vertical axis gives the percentage of atoms originating from B in the total count of atoms at fixed $x$.

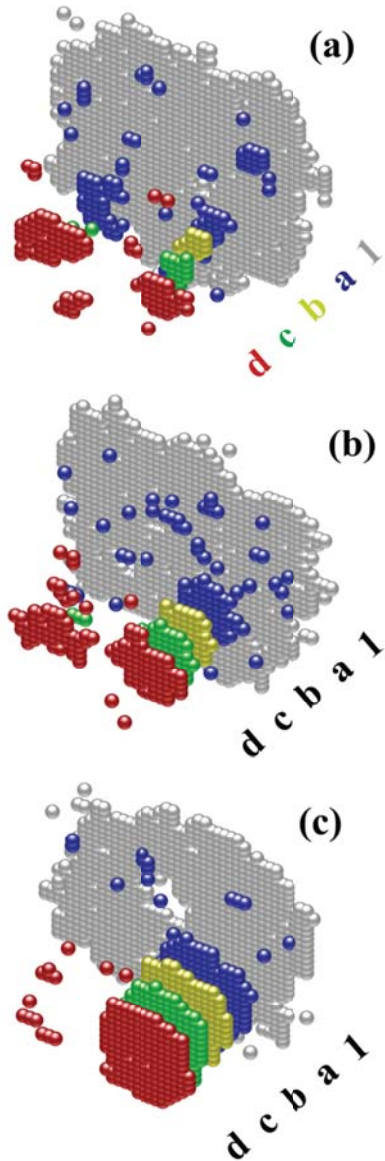

**Figure 7**. A random simulation result at a somewhat larger temperature than that in Figures 2-5, here with $\alpha'=1.8$, $p'=0.89$, showing a single realization of the neck emergence for (a) $t=3\times10^3$; (b) $4\times10^3$; and (c) $5\times10^3$. The color coding, defined in panel (a), corresponds to the layer labels, cf. Figure 1(a).

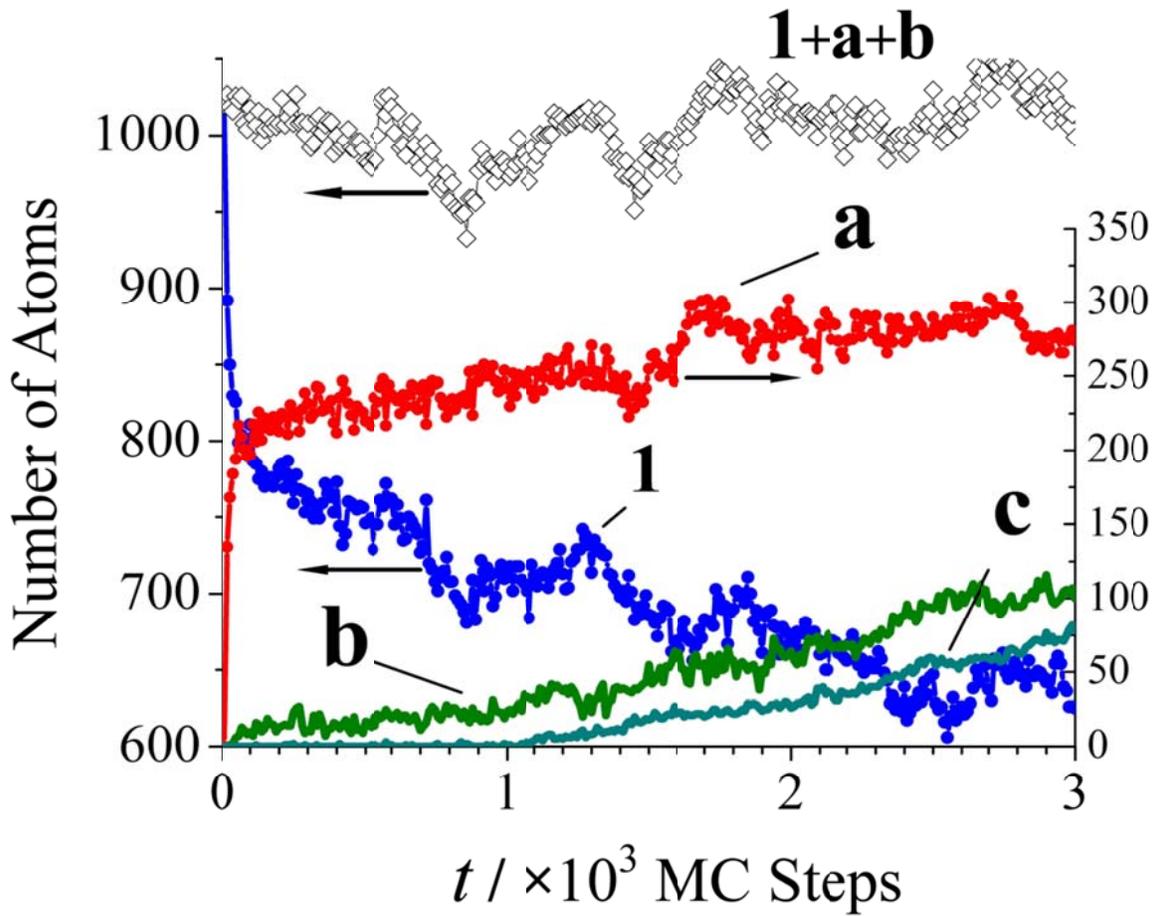

**Figure 8.** Time-dependence of the number of atoms in various layers near the gap-facing surface of particle A, in the configuration shown in Figure 1(b), for a typical MC run. For the initially empty (gap) planes **a**, **b**, **c**, the vertical axis is on the right. The left vertical axis shows the number of atoms in the initially filled layer **1**, as well as the sum of the numbers of atoms in layers **1**+**a**+**b**. The simulation parameters were $\alpha' = 2.2$, $p' = 0.87$, $R=50$, $r=40$.

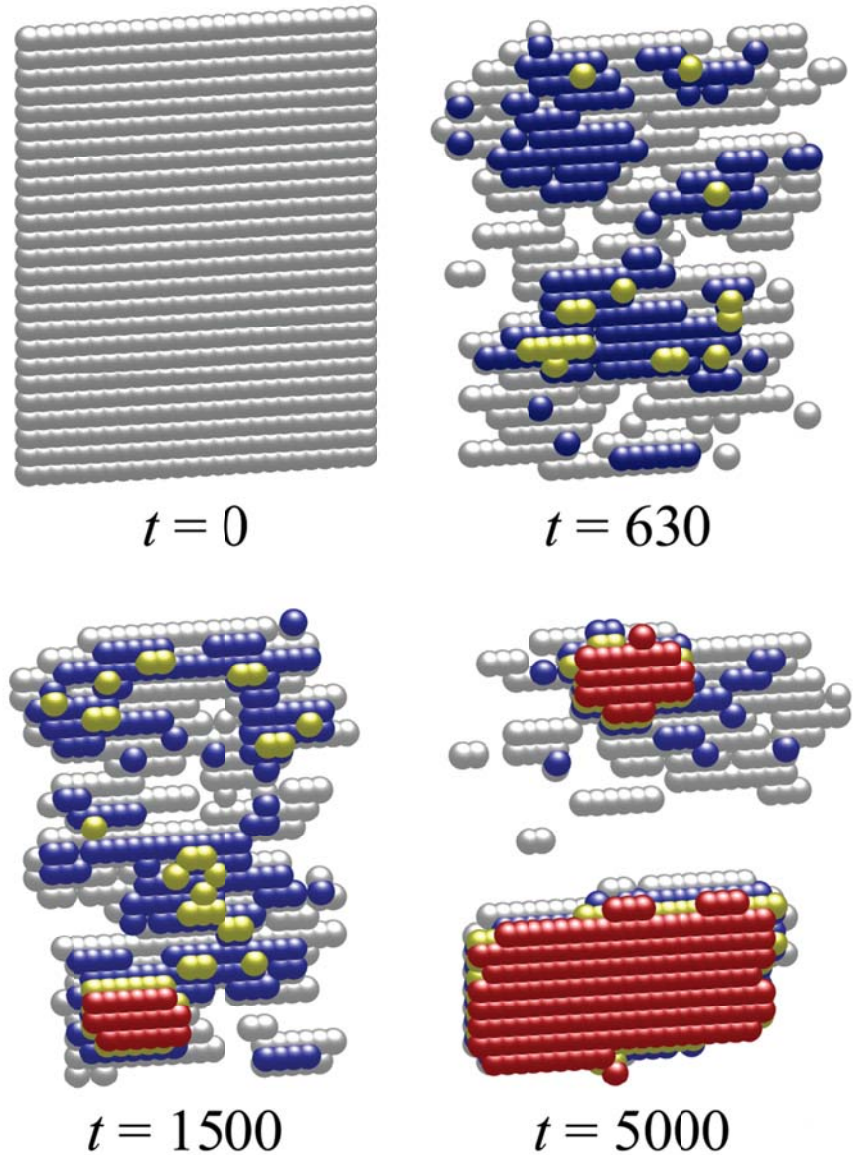

**Figure 9.** Configuration of atoms at the smaller of the two sintering (110)-type surfaces, i.e., of particle B, cf. Figures 1(b) and 8. Light gray: layer **1'**, blue: layer **f**, yellow: layer **e**, red: layer **d**. The red-marked atoms and the atoms underlying them, illustrate the formed linking structures for times $1.5 \times 10^3$ and $5.0 \times 10^3$, which are further connected via layers **c, b, a, 1,** … (not shown) to particle A.

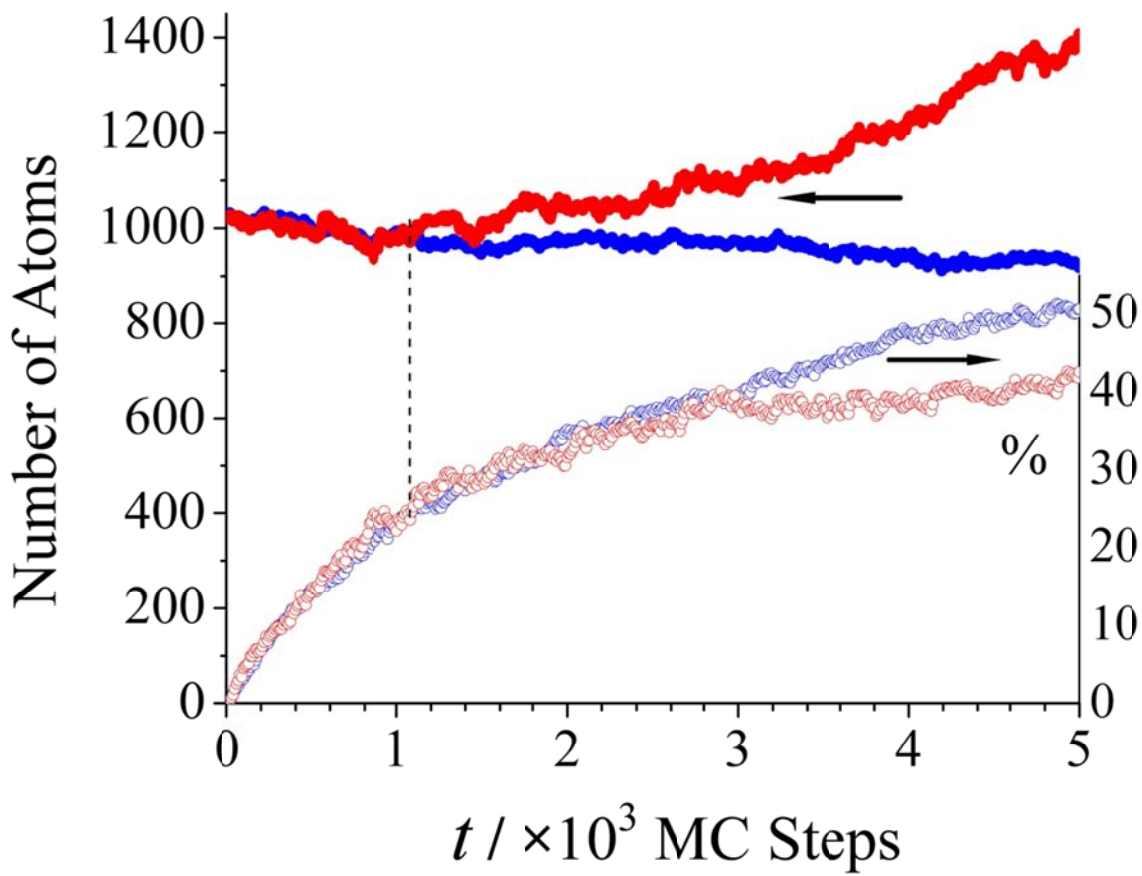

**Figure 10.** The top curves (with the vertical axis on the left) show the total number of atoms in layers **1**+**a**+**b**+**c** of particle A, as a function of time. Red: for the surface facing the gap. Blue: a similar count averaged over all those other (110)-type surfaces of particle A which are not facing particle B, see Figure 1(b). The two lower curves are similarly color coded, but give the percentages only of those atoms presently attached in layers **1**+**a**+**b**+**c** (with the vertical axis on the right) which were at any time prior to time $t$ detached, i.e., were part of the "gas" of diffusing atoms.

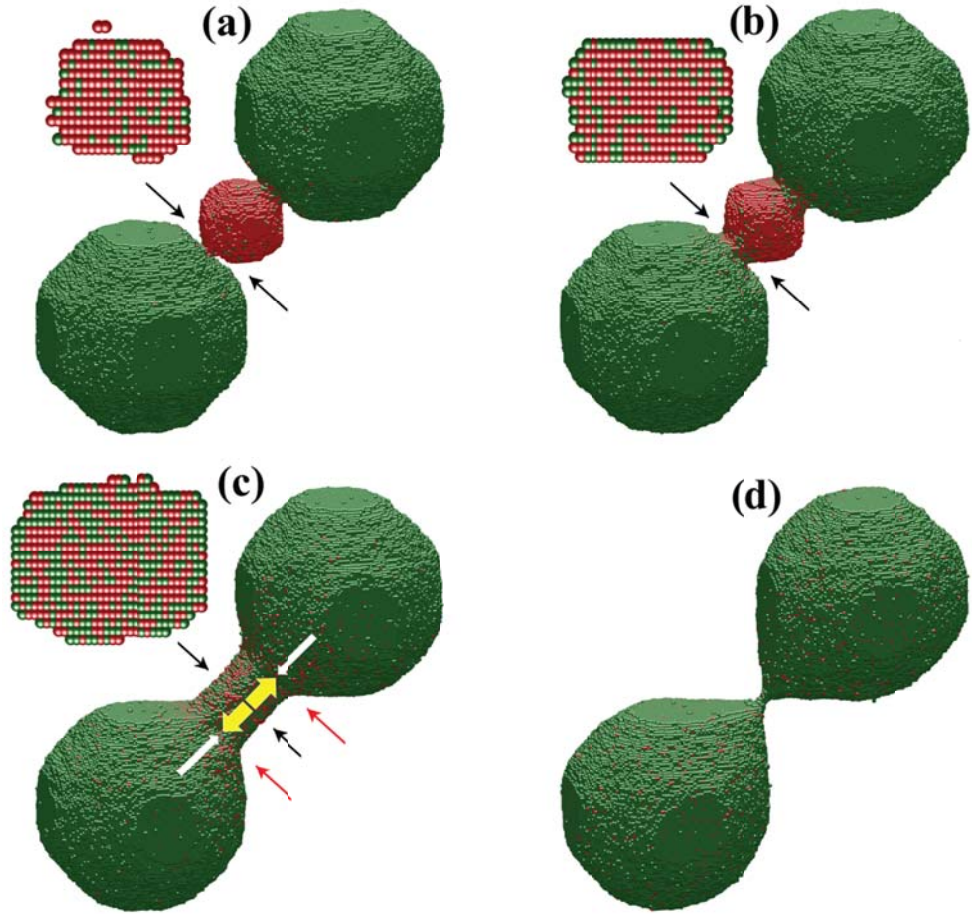

**Figure 11.** Dynamics of the configuration shown in Figure 1(b) for times (a) $t = 10^4$; (b) $2\times 10^4$, (c) $2\times 10^5$, and (d) $1.34\times 10^6$. The system parameters here were $R=50$, $r=20$, $\alpha'=2.4$, $p'=0.86$. Only the attached atoms, not in the diffuser "gas," are shown: The green-colored atoms are those which originated in particles A or C, whereas the red-colored ones are those originally in particle B. The insets in panels (a), (b) and (c) illustrate the distribution of atoms in the (110)-type cross-sections at positions marked by the black arrows. In the shown cross-sections, the ratio of the number of the "red" atoms to that of the "green" atoms was approximately (a) 11:3, (b) 11:5, and (c) 1:1. The red arrows in panel (c), mark the regions of the largest negative curvature of the connected structure's surface. Other labeling in panel (c), is described in the text.

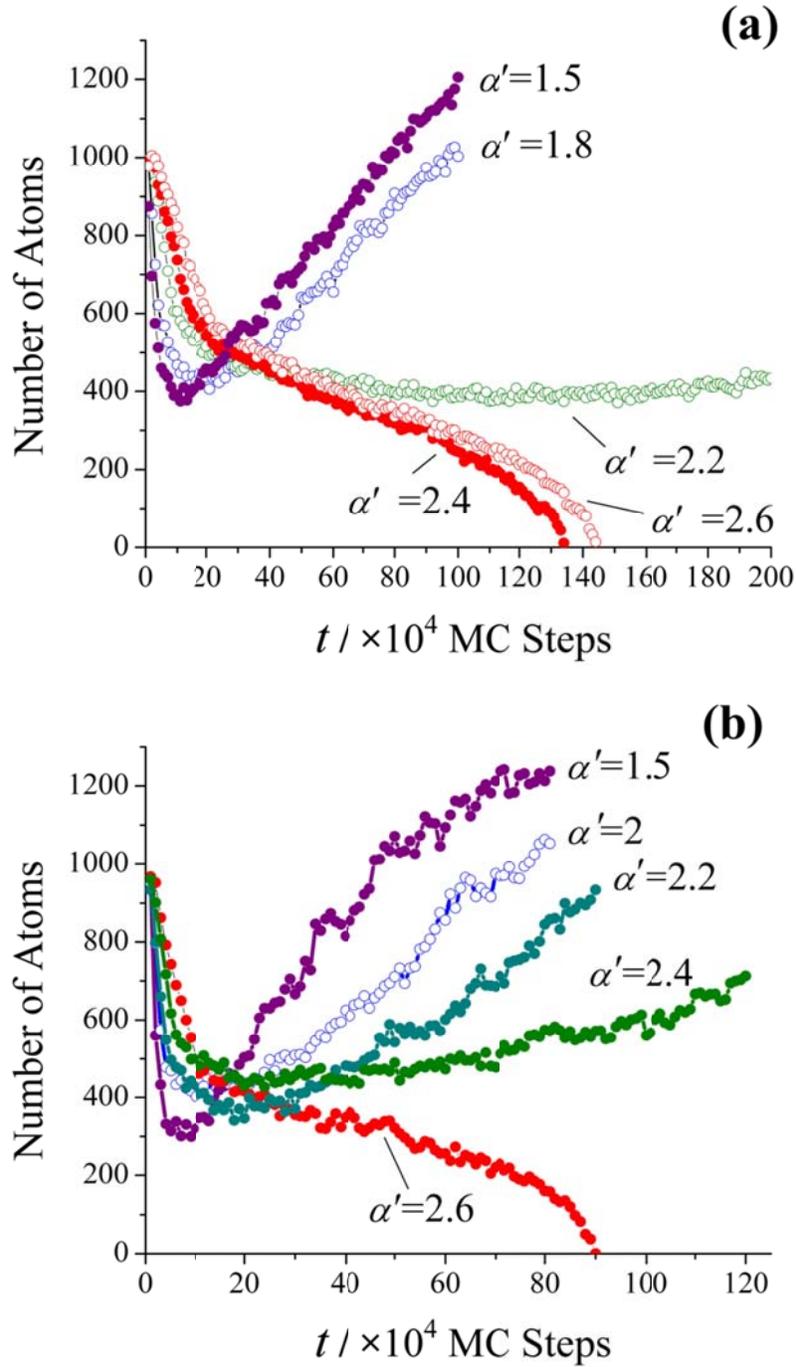

**Figure 12.** Number of atoms in the cross-sectional (110)-type plane passing through the center of particle B, in the configuration shown in Figure 1(b), as a function of time, for (a) $R=50$, $r=20$, and (b) $R=30$, $r=14$, calculated for several values of $\alpha'$.

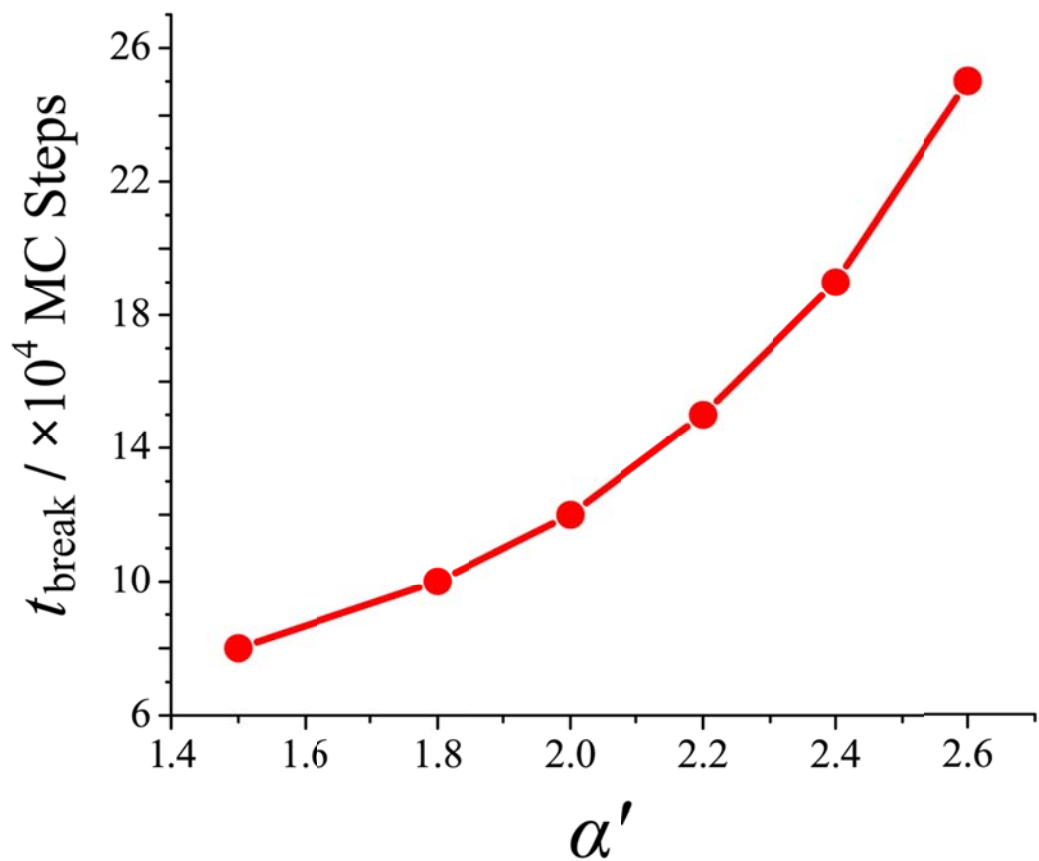

**Figure 13.** Dependence of the time it takes the connection mediated by particle B, bridging particles A and C, to break, on $\alpha'$, here calculated with the same parameters as in Figure 12(a), except that $r=14$.

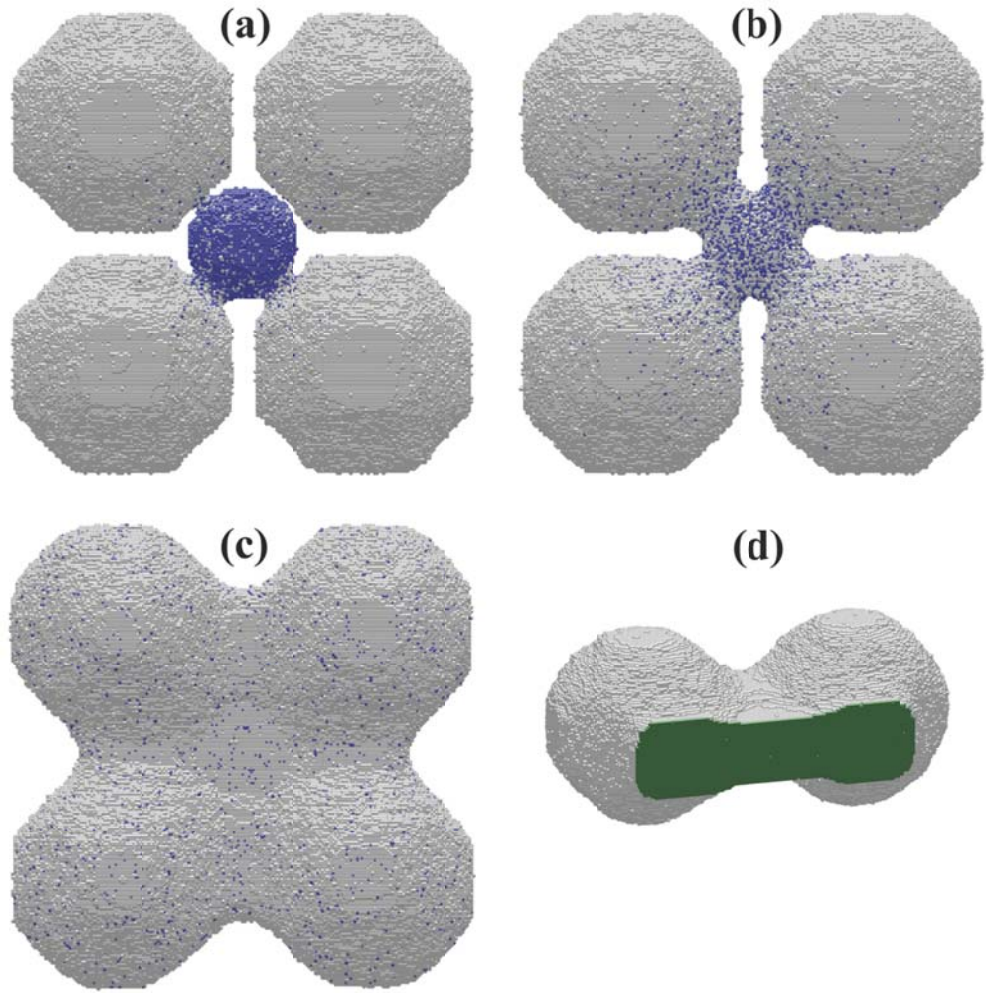

**Figure 14.** A snuggly fitting small particle inserted in the voids between four larger particles in a planar (square) arrangement, with the initial gaps the same as to those in the configuration of Figure 1(b). The parameters here, were $R=50$, $r=25$, $\alpha'=2.2$, and the gaps between the large particles were 12 lattice spacing. System evolution is shown for times (a) $t=10^4$, (b) $10^5$, and (c,d) $2\times 10^6$, where panel (d) displays the configuration (c) when symmetrically cut in half horizontally.

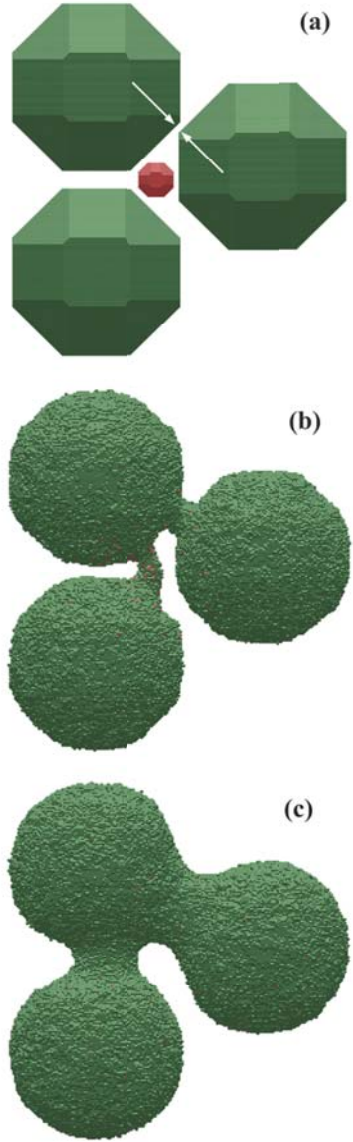

**Figure 15.** Illustration of an asymmetrical configuration of three large particles of the same size as before, $R=50$, which only allows insertion of a smaller-size small particle, $r=10$, than for earlier considered configurations. The dynamics of the systems is shown for times (a) $t = 0$, (b) $10^4$, and (c) $10^5$. Here the gaps between the small particle and each of the surrounding large particles are exactly or approximately $5\ell$. The white arrows mark the gap between the large particles which is also $5\ell$. The other two such gaps were both larger than $12\ell$. In order to ensure stable connection formation, here the temperature was set to a larger value than for the earlier considered configurations: $\alpha'=1.5$.

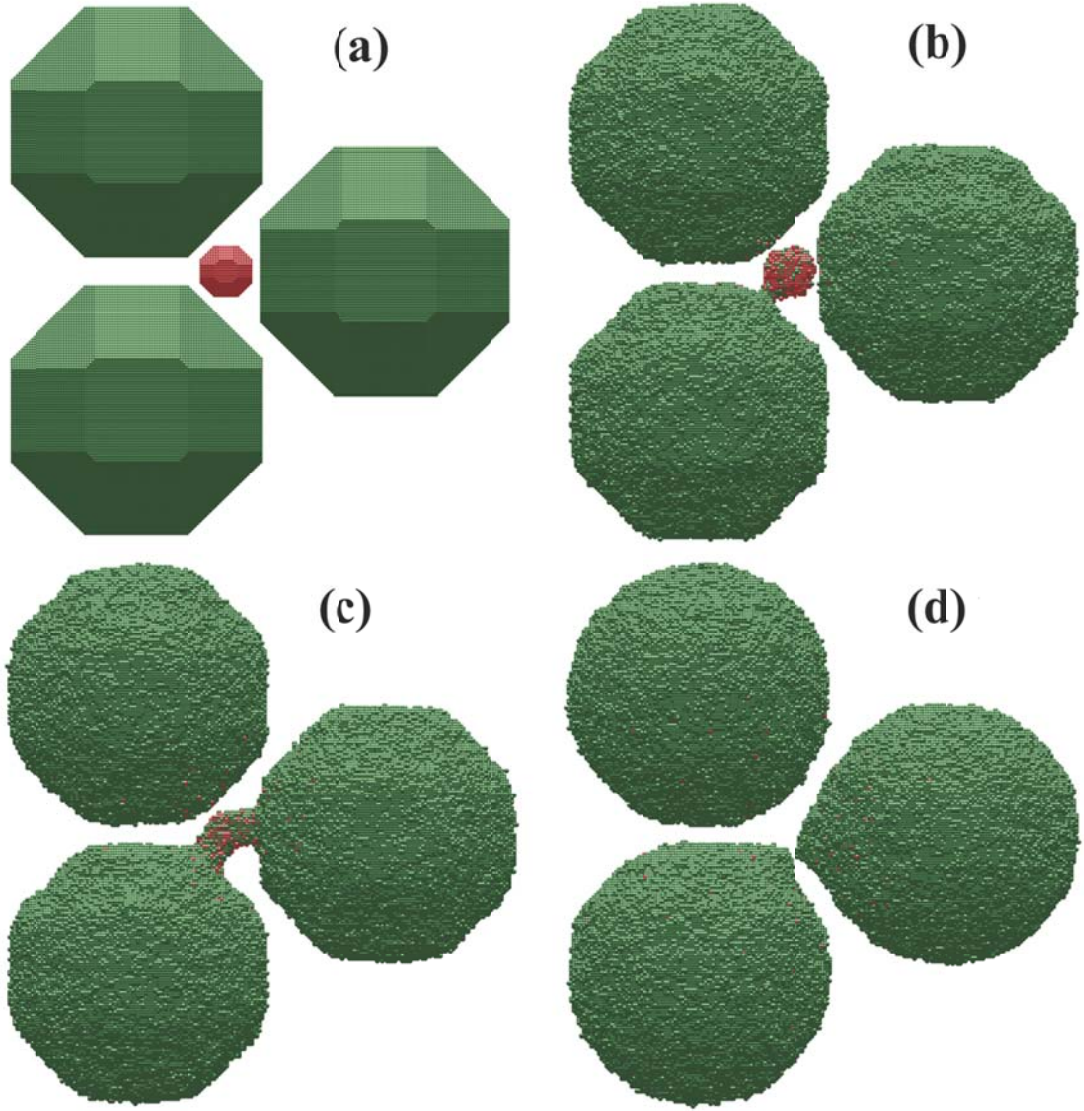

**Figure 16.** A randomly generated necking configuration resulting from the shown arrangement of particles, initially symmetrical with respect to the horizontal plane, sketched in panel (a). The system's evolution is shown here for times (a) $t=0$, (b) $2\times 10^3$, (c) $5\times 10^3$, (d) $19.8\times 10^3$. The system parameters were the same is in Figure 15, except that the gaps between the large particles, all $\geq 10\ell$, corresponded to the rightmost large particle being symmetrically positioned relative to the rest of the structure.

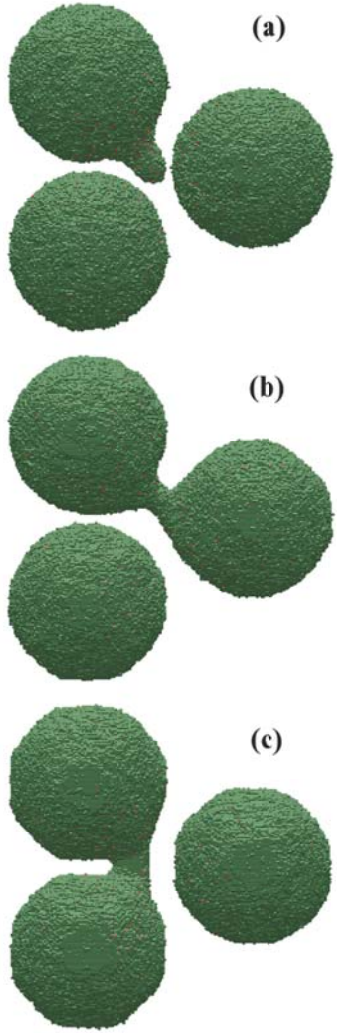

**Figure 17.** This configuration is symmetrical and similar to that in Figure 16, with a somewhat larger small-particle size, $r=14$. The larger particles, of size, $R=50$, were "fitted" around it in such a way that all their gaps with the small particle were exactly or approximately $5\ell$. This resulted in somewhat larger gaps, $\geq 12\ell$, between the large particles. The shown configurations were obtained for (a) $\alpha'=1.5$, at time $3\times10^4$; (b) $\alpha'=1.8$, at $t = 15\times10^4$; (c) $\alpha'=2.0$, at $t = 15\times10^4$.

**Table of Contents (Graphical Abstract) Entry: Image and Legend**

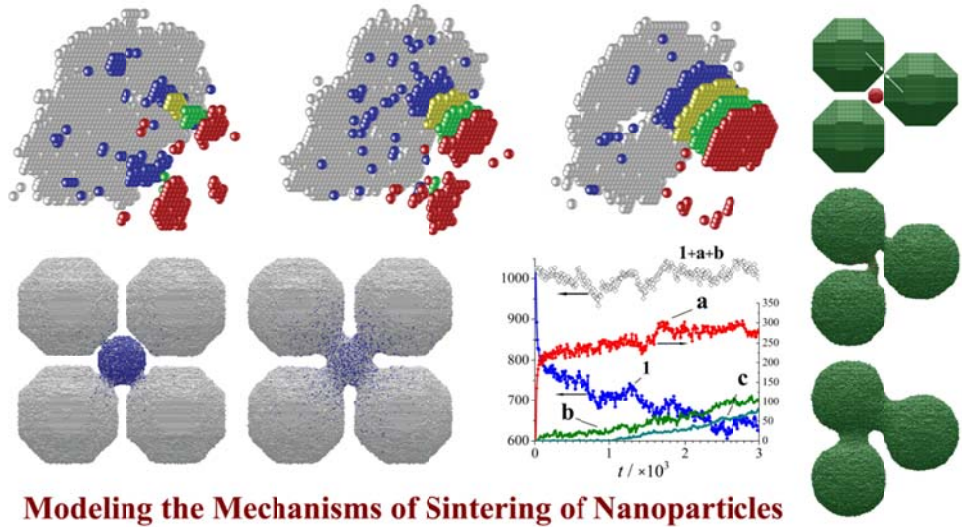

**Legend:** Sintering of dispersed nanoparticles of bimodal size distributions is modeled, including aspects of neck initiation and development, and the effect of physical and spatial-positioning parameters on particle merging.

**Journal Issue Cover**

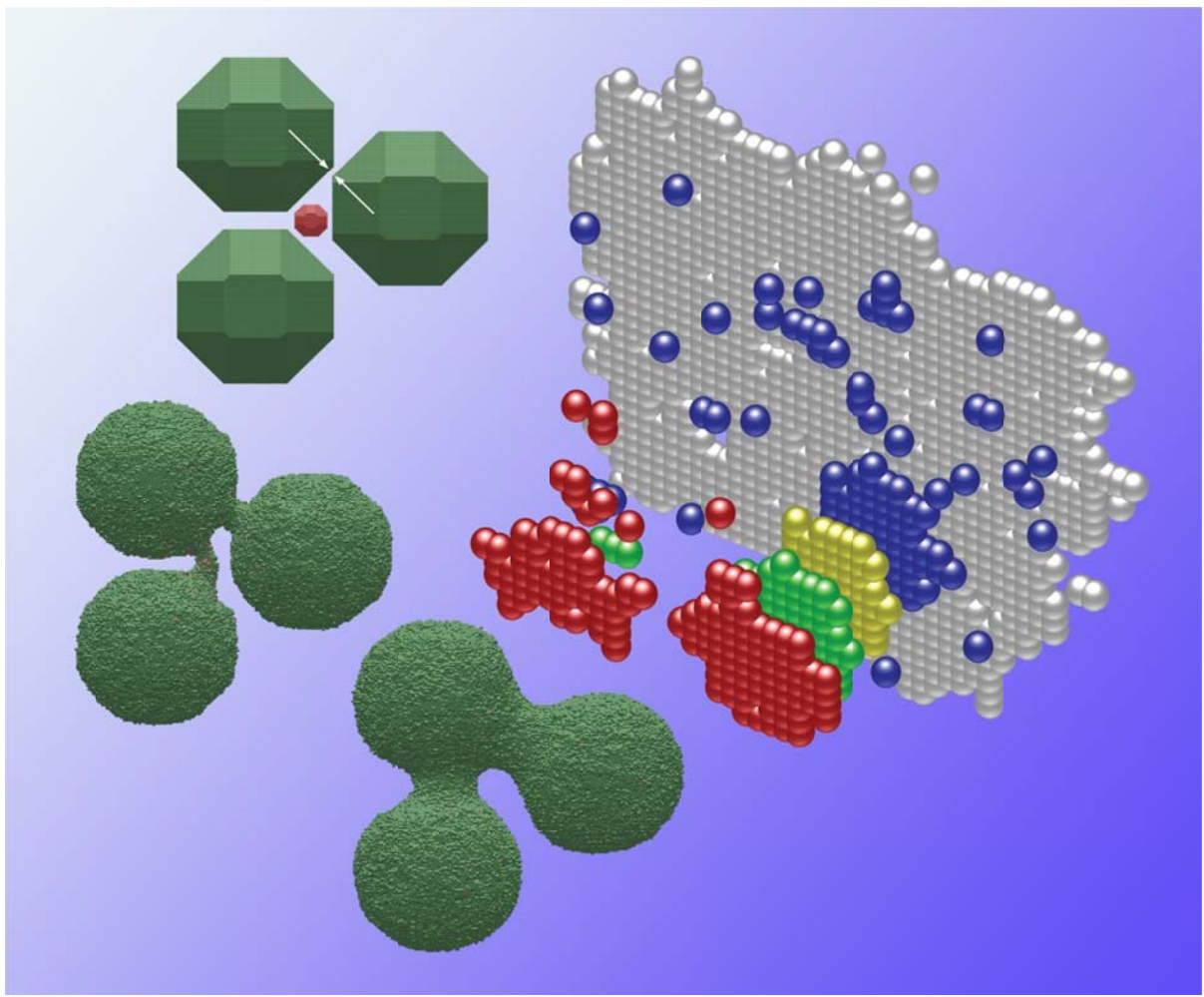

Results of this work have been highlighted by the Editors of *CrystEngComm*, by using a compilation of images for a cover of Issue 36 of Volume 15 (September 28, 2013)